# 1. DESIGN OF BEAM OPTICS AND RADIATION PROTECTION CONCEPT FOR NA60+ HEAVY-ION EXPERIMENT AT CERN

A. Gerbershagen† [1,2], C. Ahdida[1], J. Bernhard[1], V. Clerc[1], S. Girod[1], E. Scomparin[3], G. Usai[4], H. Vincke[1]

[1] European Organisation for Nuclear Research (CERN), Meyrin, Switzerland

[2] PARTREC, UMCG, University of Groningen, The Netherlands

[3] INFN Sezione di Torino, Turin, Italy

[4] Dipartimento di Fisica dell' Universita' di Cagliari and INFN Sezione di Cagliari, Cagliari, Italy

*Abstract*

NA60+ is a fixed target experiment proposed in the framework of the Physics Beyond Colliders programme at CERN. It aims to precisely measure the hard and electromagnetic probes in nuclear collisions. Initially proposed for the underground cavern ECN3 with very high beam intensities, the experiment now foresees a location in the EHN1 surface hall which was shown to have a limited impact on the physics performance in spite of a significant reduction of beam intensity and detector size. The potential installation and operation of the experiment with the ion beams from the Super Proton Synchrotron (SPS) has been examined regarding detector integration, beam physics, radiation protection and shielding requirements. The integration of the experiment is considered feasible, but would require a significant reconfiguration of the existing hall infrastructure with regards to shielding and layout.

## INTRODUCTION

*Physics Beyond Colliders*

Physics Beyond Colliders (PBC) [1] is a study initiated at CERN in 2016 to explore the possibilities of fundamental physics research complementary to that at existing or future colliders. A considerable share of the proposed programme focusses on fixed target experiments to be located in CERN North Experimental Area [2] supplied by beam from the CERN Super Proton Synchrotron (SPS), as depicted in Figure 1.

---

† a.ge@cern.ch

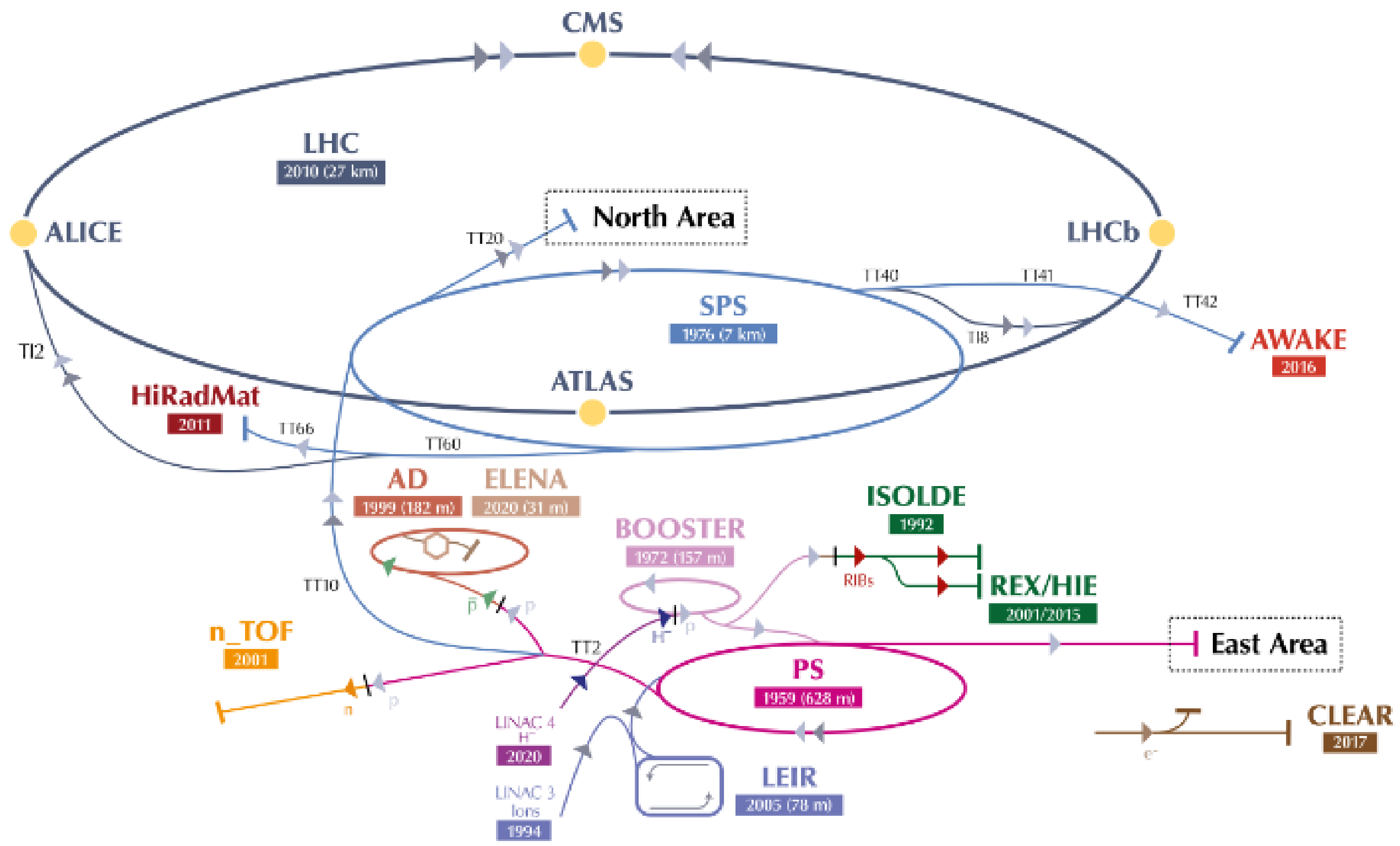

Figure 1: CERN Accelerator Scheme, with the North Area within the LHC circle in the center of the scheme [3].

Many experimental proposals have been analysed by the PBC Conventional Beams Working Group [4]. The present technical note focusses on the evaluation of the feasibility of the integration of the NA60+ experiment from the mechanical, topological, beam physics and radiation protection points of view. It does not cover the physics scope of the experiment or the feasibility or performance of its detectors.

*NA60+ Experiment*

The NA60+ experiment [5] is proposed to study the production of thermal dimuons and open charm using a primary ion beams from the SPS in the momentum range of 30-158 A GeV/c.

The current schematic layout of the detector is shown in Figure 2. It consists of a target located inside a large normal-conducting dipole magnet followed by a silicon vertex spectrometer, an absorber, a muon spectrometer combining a normal conducting toroidal magnet with five tracking stations and a muon wall. The transverse size of the detector has been reduced from the initial value proposed of 9.0 m in [5] to 6.2 m, reducing in this way its angular acceptance but without significantly affecting the overall physics performance. The length of the detector can vary between 10.4 m for low momenta to 13.7 m for the higher momenta, for which the downstream part of the detector, consisting of the detector planes, toroidal magnet and the muon wall, needs to be movable (see a possible integration solution in Section 3 of the present article).

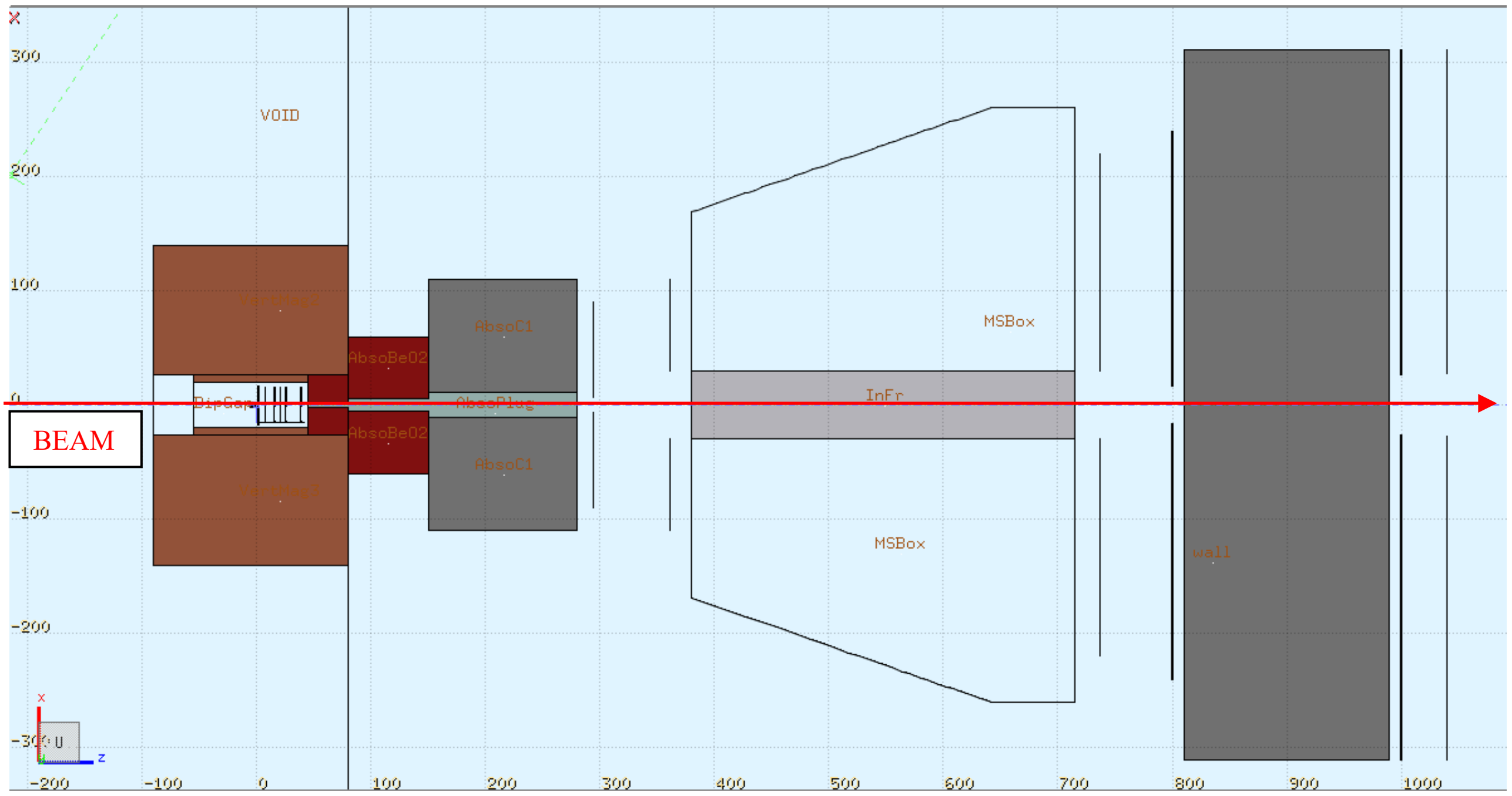

Figure 2: Visualisation of the FLUKA [6] [7] model of the NA60++ experiment using FLAIR [8]. Brown color represents the yoke of the magnet around the target, which is located in air. The absorber material is depicted in red and grey inside of the magnet and in the dedicated absorber module (indicated as Absoplug, AbsoC1 and AbsoBe02), followed by toroidal magnet (colourless) amd muon wall (grey). The scale both in X and in Y is in cm.

*Current EHN1 hall layout*

CERN North Area receives primary or secondary proton or ion beams from the SPS accelerator. The primary SPS beam (arriving from the top left in Figure 3) is being split into three branches, each one transporting a fraction of the SPS beam intensity towards the three North Area Targets (“T2”, “T4” or “T6”), where secondary beams are created. The attenuated primary beams can also be transported downstream to the experimental halls via the secondary beam lines “H2”, “H4” and “H8” and to the three experimental halls “EHN1”, “EHN2” and “ECN3”.

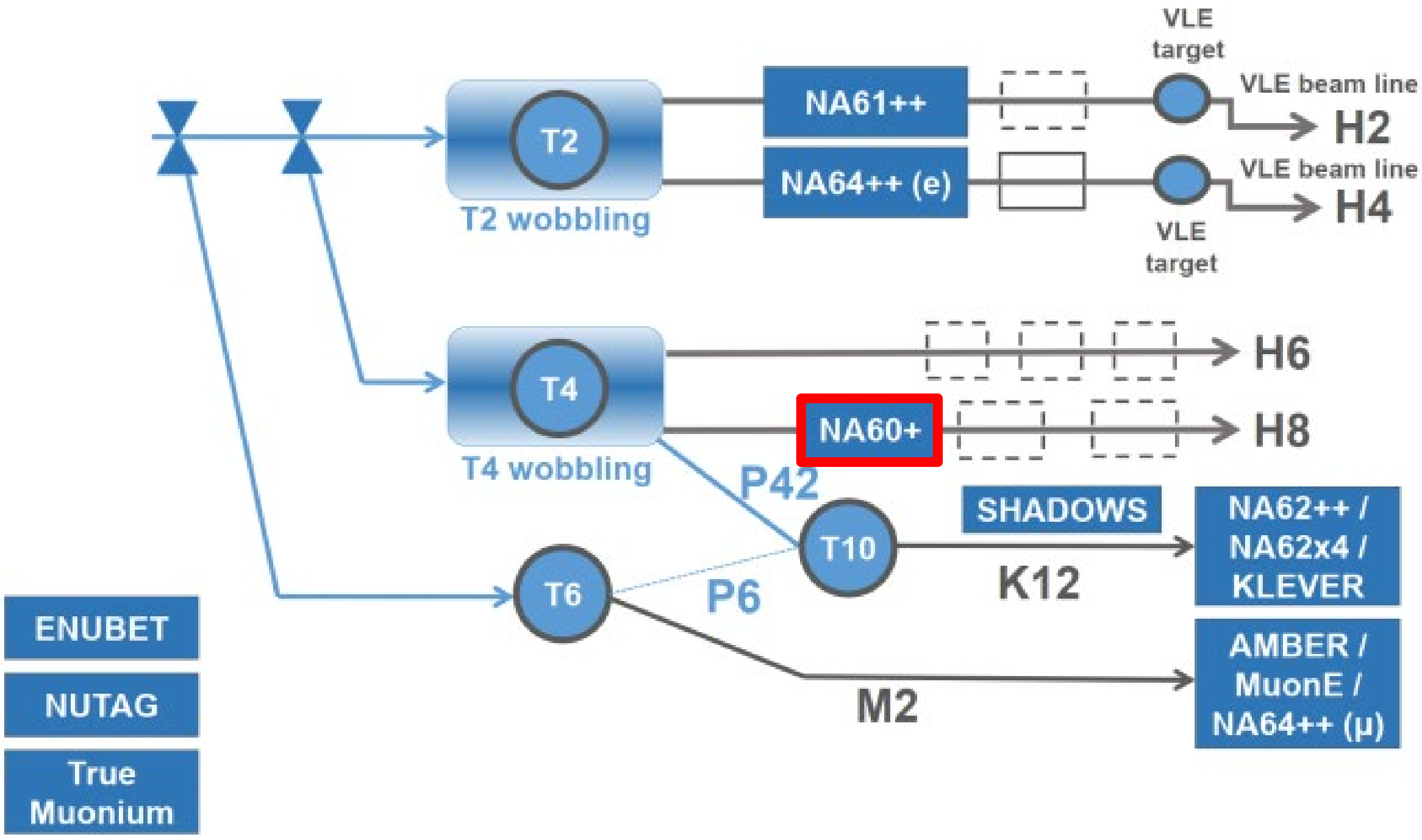

Figure 3: Schematic diagram of the North Area beam lines and the experiments proposed in the framework of the Physics Beyond Colliders programme. The dotted boxes without names are experimental zones for the usual test beam users in the area. ENUBET, NUTAG and True Muonium are site-independent proposals. The present note is focusing on the NA60+ experiment (marked with red box), with a proposed location at H8 beam line of EHN1

Several locations have been considered for accommodating the NA60+ detector in the North Experimental Area. These include an initially proposed location in the underground ECN3 cavern for high intensities, and several zones of the EHN1 surface hall, including the PPE138 zone of the H8 beam line. Locations in the EHN1 surface hall can only be considered for lower beam fluxes of up to $10^7$ Pb ions per spill. Figure 4 shows the layout of the EHN1 hall and indicates the user zones within the hall. The four beamlines H2, H4, H6 and H8 transverse the hall from the left towards the right side of the diagram. Their user zones are marked by green, blue, violet and red colours, respectively.

Zone PPE138 (in the bottom left quarter of Figure 4) was identified to be the most promising location, considering the absence of other major fixed target experiments in this beamline (competing for space and beam time), and the full spectrum of requests within the PBC programme for the other zones.

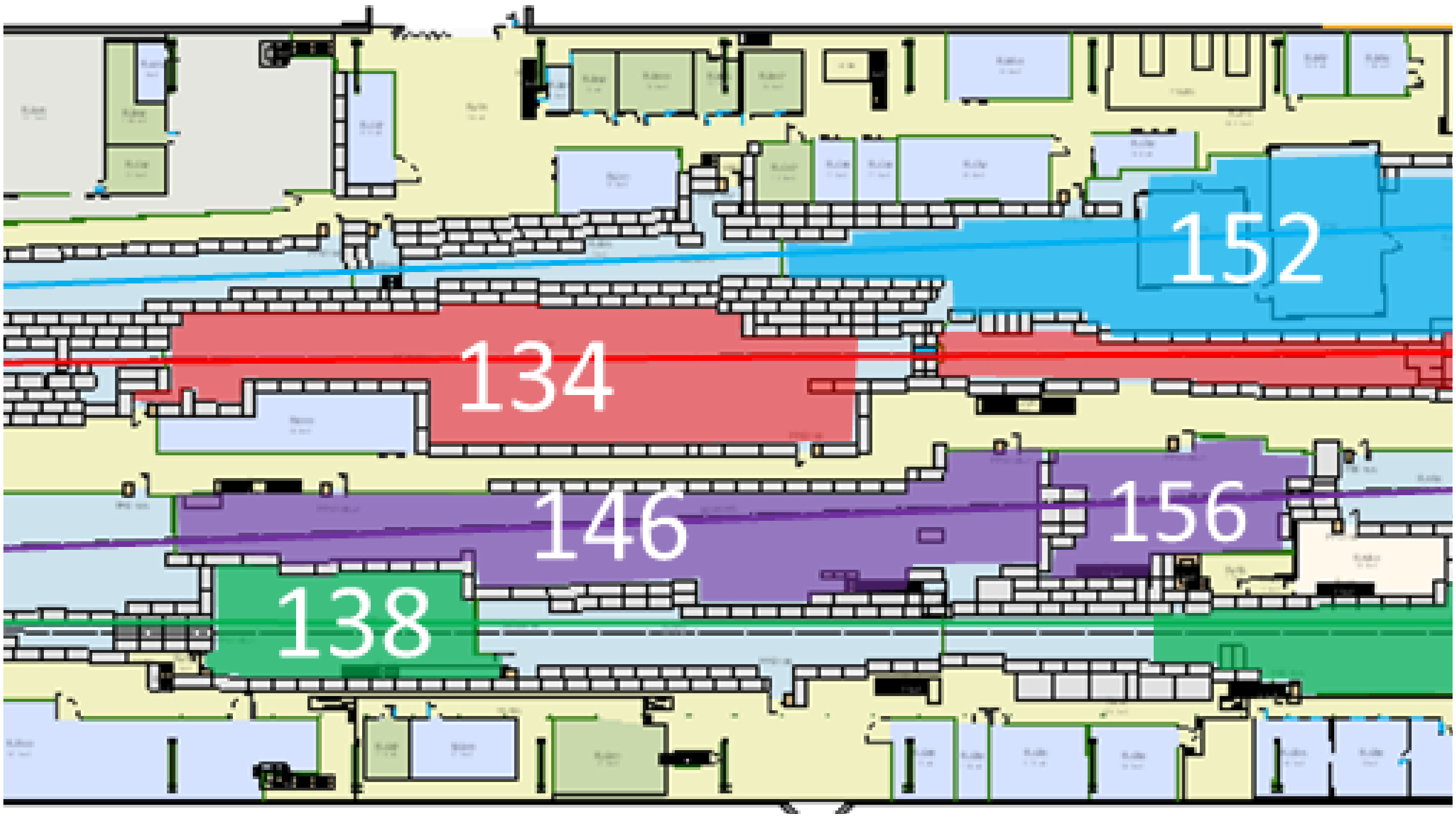

Figure 4: Layout of the upstream part of EHN1 hall with its user zones. The counting rooms next to the user zones are indicated as boxes along the top and the bottom of the scheme.

The beam enters the zone from the left side and travels towards the right. In order to accommodate the experiment, the zone would need to be substantially modified with regards to its shielding, access and layout. The reasons behind these modifications and the details of the modification will be described in the following sections.

## PROPOSED BEAM SET-UP

The slow extraction of the ions from the SPS into the CERN North Area is performed through debunched spills of about 9 s duration. The duty cycle of the SPS extraction to the North Area depends on the energy of the extracted beam, since higher energies require higher current in the SPS magnets, and their cooling efficiency is one of the limiting factors for the duty cycle. The other factors include the parallel programs of SPS use, such as LHC injection with protons or ions, Machine Development or degaussing magnetic cycles, etc. The maximal duty cycle can be calculated to be around 40%, with approx. 9 s spills within an approx. 25s sypercycle. The duty cycle value of 50% has been taken as a baseline for the investigation of Radiation Protection (RP) related issues, beam optics calculation and the integration design. This value provides a reasonable margin to the currently achievable maximum of 42% and might become conceivable in case of major upgrades during the proposed operational period of NA60+ experiment (until around 2040).

The beam intensity required to fulfil the NA60+ physics programme in EHN1 is $10^7$ primary lead ions per spill at the target of NA60+, which corresponds to $5{\cdot}10^5$ ions per second (or $10^6$ ions per second with 50 % duty cycle). The beam spot size at the experiment needs to be as small as possible, with the beam fitting within a 4 mm hole in the central part of the detector.

The requested beam intensity can routinely be delivered by the accelerator chain and strong collimation will be needed to reduce the intensity delivered to $10^7$ ions per spill at the experiment, where the intensity is limited by the RP considerations. The beam parameters at the beginning of the H8 beam line are not well known due to the lack of precise beam instrumentation. At the T2 target, where the beam conditions are not identical, but comparable, a measurement of the beam size was performed in 2017, showing that the 150A GeV/c lead ion beam had the profile displayed in Figure 5 with an overall beam size of approximately 1x1 mm.

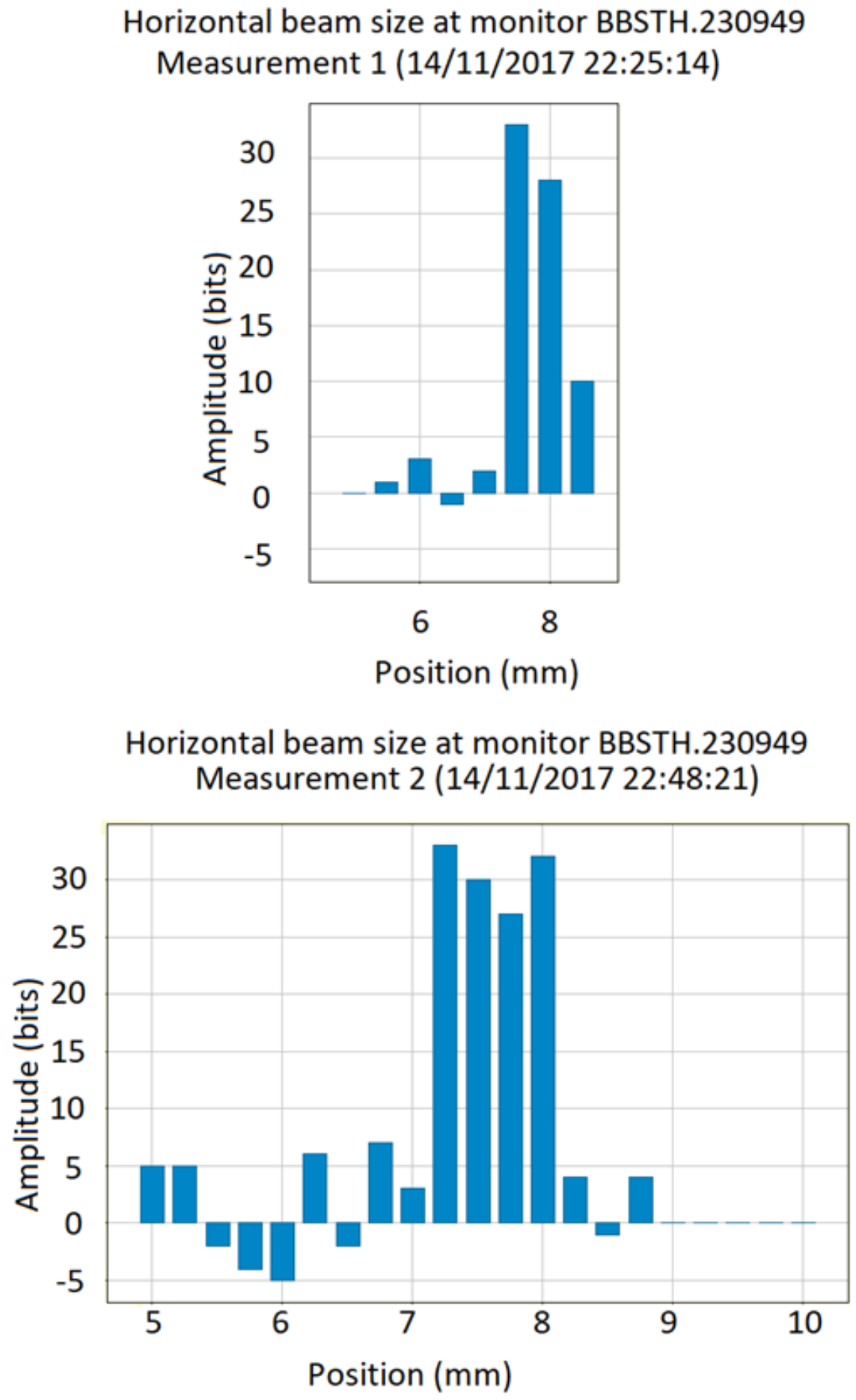

Figure 5: Two different measurements using the motorized secondary electron emission monitors installed in front of the T2 target in TCC2. The horizontal beam size of the beam is in both measurements of the order of a few microns RMS [9].

Based on this measurement and for the purpose of this study, a conservative assumption has been made that the 150A GeV/c lead ion beam at the T4 location would have an RMS beam size of 0.5 mm (full size of ~2 mm) and that the beam divergence has a comparably large RMS value of 0.5 mrad (see Table 1). In order to estimate the values for a low energy beam of 30A GeV/c, only the geometrical change of divergence (proportional to $1/\sqrt{p}$) has been considered. In reality additional changes of beam size and divergence can be expected due to the limited precision of SPS rectifiers and the reduced response from the beam instrumentation at lower momenta, which can impede the beam steering and aggravate the beam losses. However, the exact amount of their contribution is difficult to estimate. It should also be noted that the RMS of the initial beam divergence is a less important parameter for the estimation of the beam size, since the maximal divergence is limited by the H8 beamline acceptance, which itself is dependent on the apertures and optics settings of the beamline. The initial divergence is, however, relevant for the estimation of the relative transmission through the H8 beamline for the different beam optics options.

Table 1: Assumptions of initial lead ion beam parameters at the start of the H8 beamline

| **Parameter** | **160A GeV/c** | **30A GeV/c** |
|---|---|---|
| $\sigma_x$ , $\sigma_y$ (mm) | 0.5 | 1.15 |
| $\sigma_{px}$, $\sigma_{py}$ (mrad) | 0.5 | 0.5 |
| $\sigma_p$/p (%) | 0.1 | 0.1 |

The current beam optics settings used in the ion operation of H8 line would deliver a beam with an RMS transverse size of 0.8 mm at the location of the experiment, which is larger than requested by NA60+. Hence, two beam optics settings have been developed, with the aim of reducing the beam size. Both optics versions are achromatic to first order. One is an existing optics, based on the use of the so-called “Microcollimator” – a very small and precisely aligned collimator used for the primary proton beam operation in H8 (see Figure 6). The Microllimator optics [10] have been routinely used in proton beams involving crystal channelling studies. It provides high beam stability, since the beam is imaged from the well-defined physical gap of the Microcollimator to the location of the experiment. In the first approximation, each of the two Microcollimator gap opening settings is linearly proportional to the beam transmission through the collimator and to the beam size in the respective plane at the location of NA60+ target.

However, Microcollimator optics has never been utilized with the ion beam, and it is not yet experimentally tested how well the ion beam can be focussed at the Microcollimator. Insufficient focussing would lead to inferior transmission, which is determined by the beam size at the Microcollimator and the Microcollimator aperture. In addition, it is not known what amount of fragmented ions would propagate further to the experiment, hence creating the background noise, generating unnecessary detector, beam line and tunnel activation and contributing to the dose at the RP monitors. These aspects need to be further studied and experimentally tested in more detail.

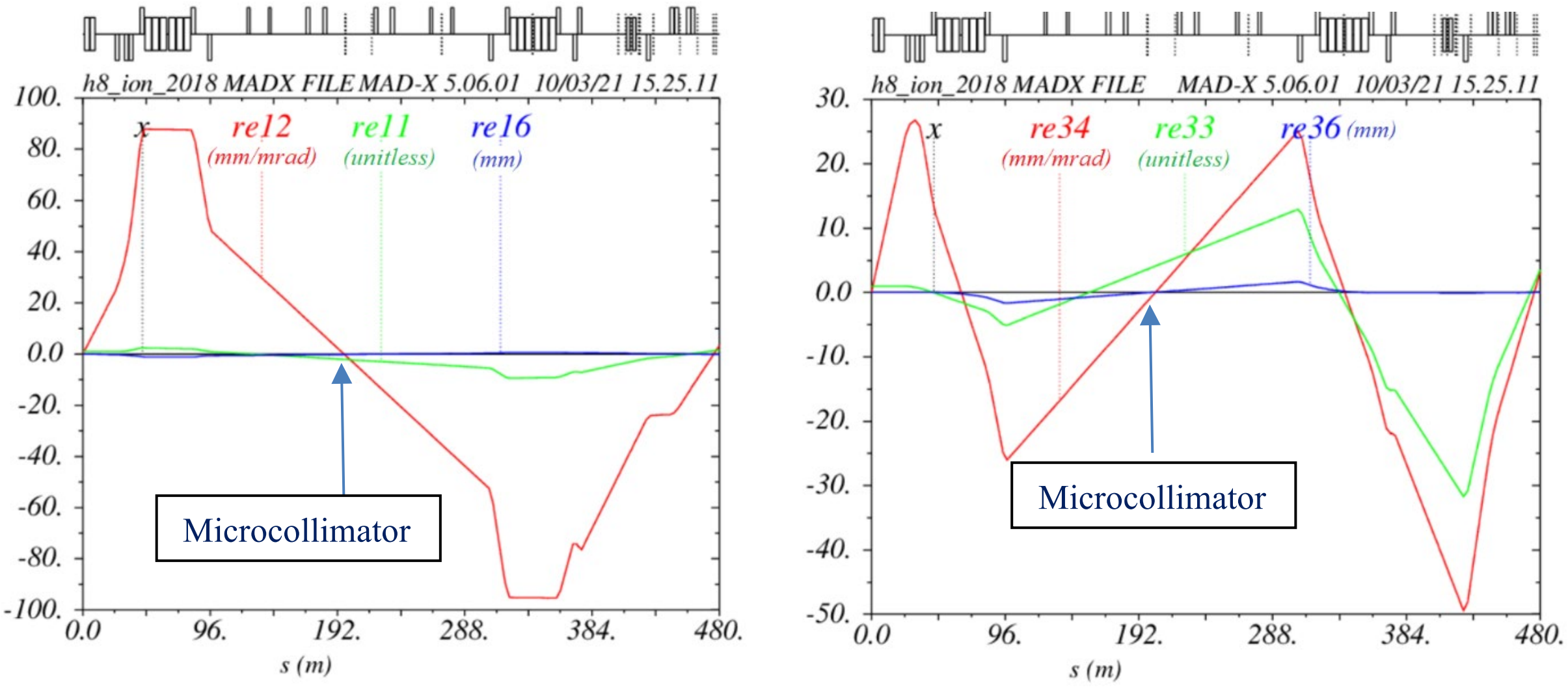

Figure 6: Optical transfer matrix functions (cosine-like in green, sine-like in red and dispersion in blue) for the horizontal (left) and vertical (right) plane for the Microcollimator ion beam optics for NA60+. Horizontal axis is the position in meters along the H8 beamline.

The second optics does not use the Microcollimator, but instead relies on stronger focussing of the beam at the experiment location (see Figure 7). It has the large advantage of allowing much higher transmission of the beam to the experiment location without compromising the beam size and is hence currently the preferred option.

The resulting beam size at the experiment location and transmission through the H8 beamline are summarized in Table 2. It should be noted that these results are based on optimistic assumptions about the beam size at the target and ignore the effects of beam transport and scattering in air and in the beam line diagnostics elements. The tracking has been performed with MADX PTC and ignores all interactions with material in the beam line and production of secondary particles.

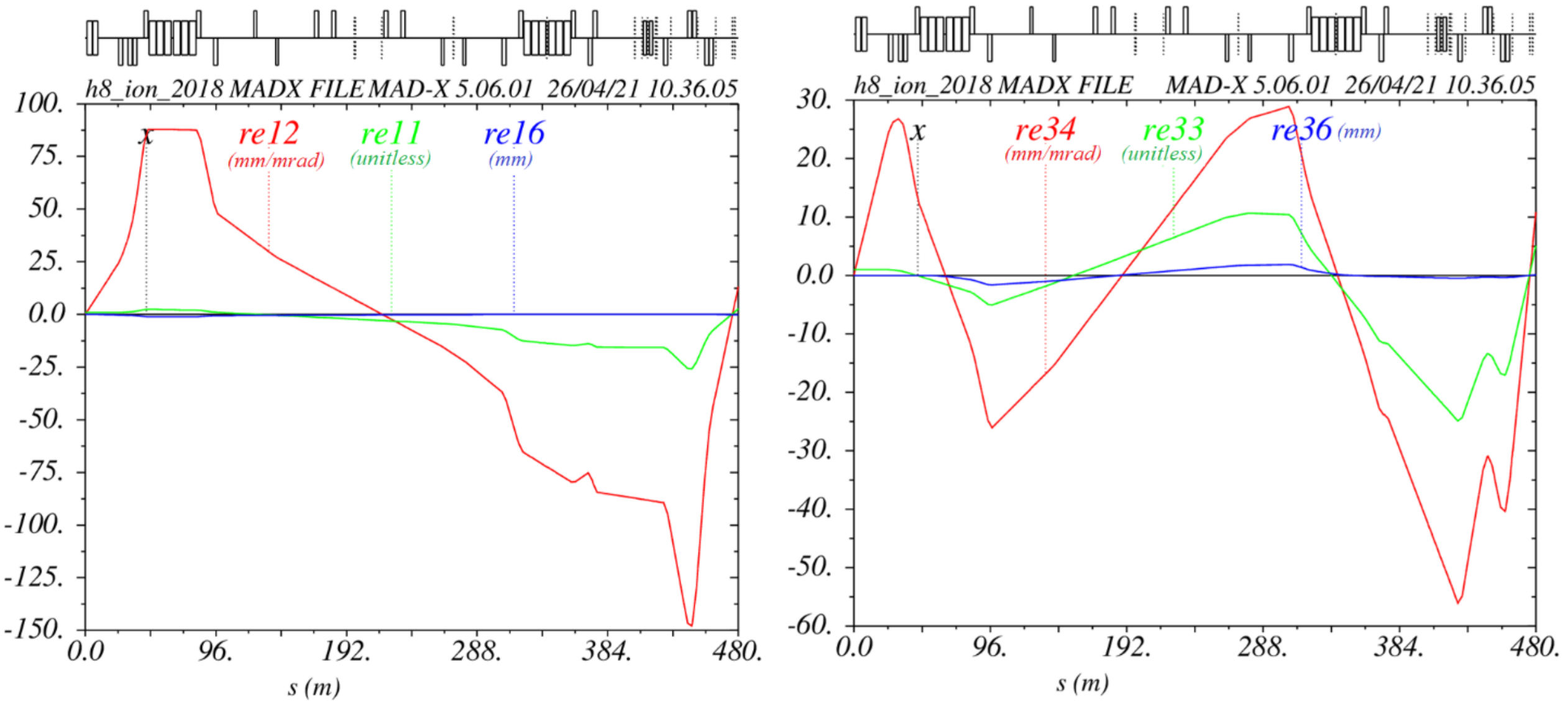

Figure 7: Optical transfer matrix functions (cosine-like in green, sine-like in red and dispersion in blue) for the horizontal (left) and vertical (right) plane for the new, strongly focussed ion beam optics for NA60+. Horizontal axis is the position in meters along the H8 beamline.

Table 2 a: Summary of beam parameters at the potential location of NA60+ for Microcollimator design

| **Parameter** | **160A GeV/c** | **30A GeV/c** |
|---|---|---|
| $\sigma_x$ (mm) | 0.33 | 0.35 |
| $\sigma_y$ (mm) | 0.34 | 0.36 |
| Transmission from the start of the H8 beamline (%) | 12.22 | 2.91 |
| Beam spot scatter plot | 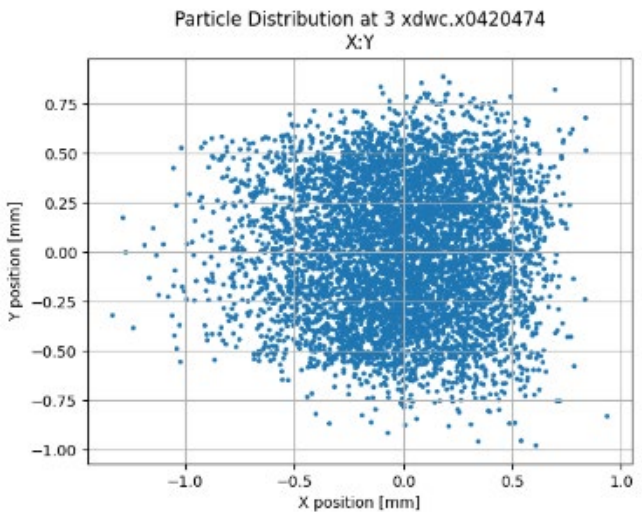  | 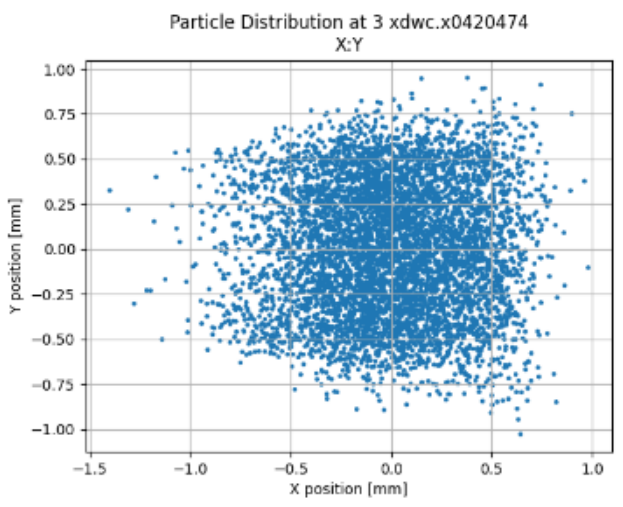  |

Table 2 b: Summary of beam parameters at the potential location of NA60+ for the focusing optics design

| **Parameter** | **160A GeV/c** | **30A GeV/c** |
|---|---|---|
| $\sigma_x$ (mm) | 0.19 | 0.33 |
| $\sigma_y$ (mm) | 0.19 | 0.36 |
| Transmission from the start of the H8 beam-line (%) | 32.43 | 23.5 |
| Beam spot scatter plot | 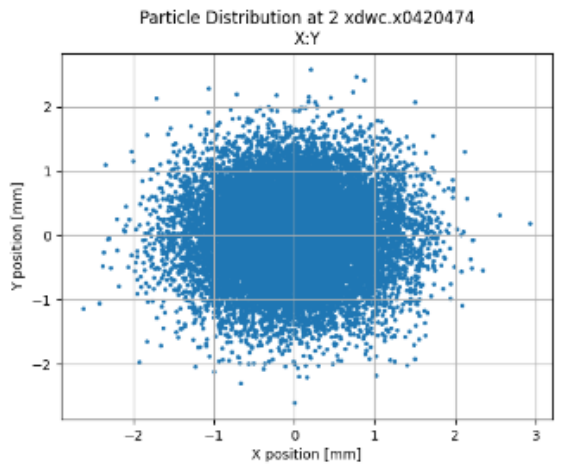 | 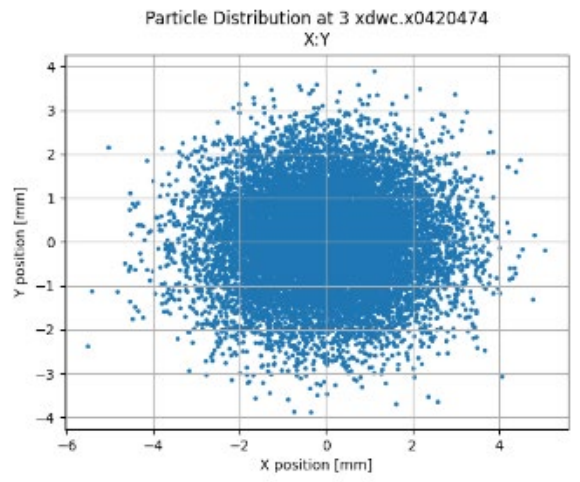 |

## ZONE LAYOUT AND INTEGRATION STUDIES

Currently, the experimental zone foreseen for NA60+ installation is used to provide test beams to several users and is therefore not optimally laid out for the installation of a major detector. An integration study has been conducted, revealing that the placement of the newly proposed reduced size NA60+ detector (radius of 3.1 m and maximal length of 13.7 m) is feasible, provided that the zone is substantially modified. The 3D-drawing of the modified design is displayed in Figure 11.

The shielding wall separating the experimental area from the user zones of the H6 beam line, will have to be moved by 80 cm towards the H6 beam line, which does not present a major problem. A new bridge and stairs to access the detector will also need to be installed (depicted green in Figure 8). Since the H8 beam height above the hall ground is 2.88 m, an excavation needs to be performed for accommodating a detector with 3.1 m radius and its mounting structure. The depth of the excavation has been set to 1 m and the transverse dimension to 6 m. There exist technical galleries The longitudinal extent of the excavation needs to cover the two setup lengths, the short one (10.4 m total length, see Figure 8) for the low energy run and the full length one (13.7 m total length, see Figure 9) for the high energy run. Rails must be installed on the bottom of the excavated area to enable the longitudinal movement of the toroidal magnet and of muon wall for the modification of the setup between the short and the long version. There exist transverse and longitudinal underground gallieries below the ground of EHN1 hall- (see Figure 10). The proposed excavation has a sufficient distance to those galleries and hence the excavation will not interfere with the structural stability of the hall and of the installed experiment. The underground galleries are closed during the beam operation, hence the Radiation Protection aspects related to prompt radiation have not been examined for those galleries.

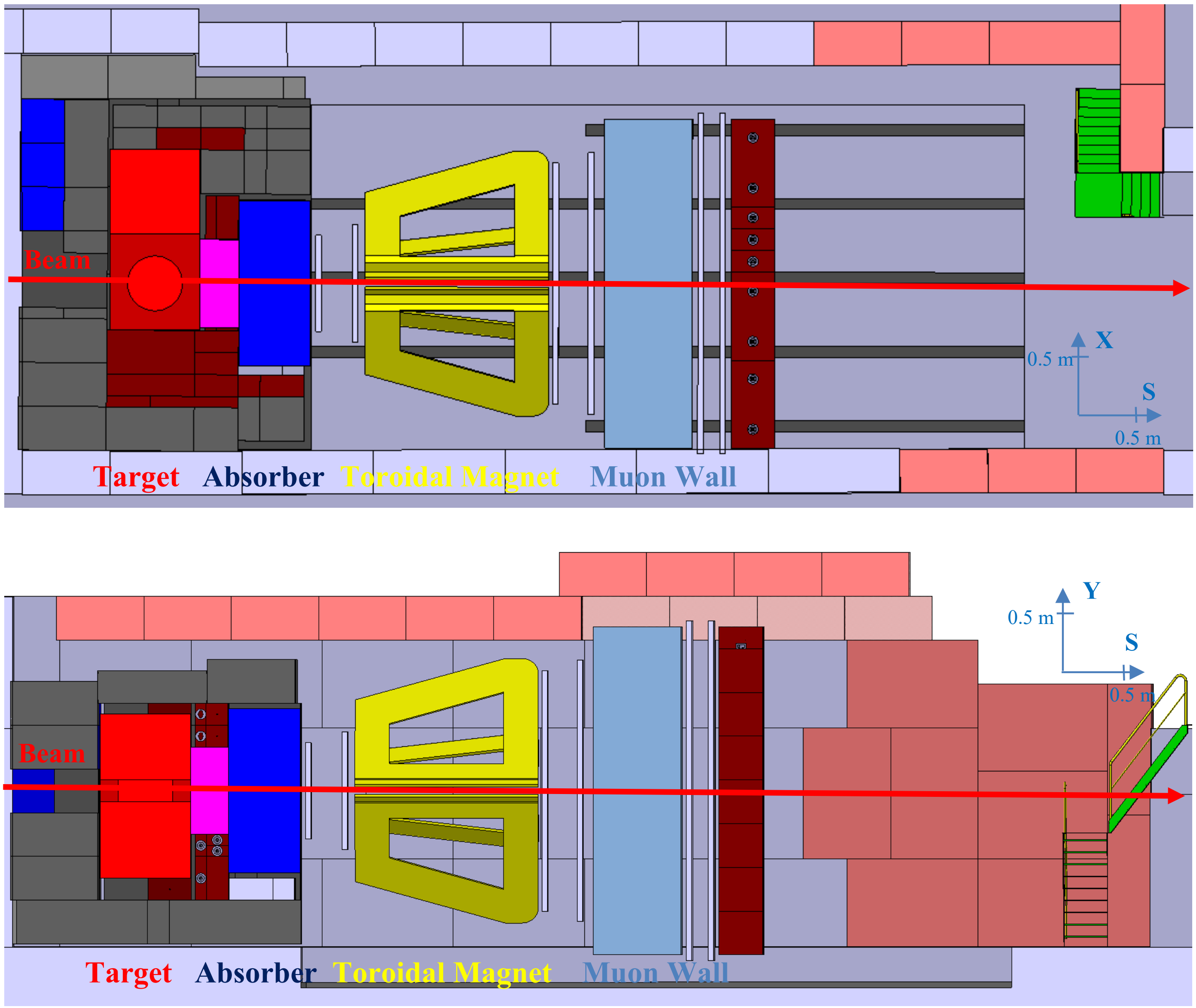

Figure 8: Drawing of the short (10.4 m) setup of NA60+ installed in the modified PPE1138 zone, top view (top) and side view (bottom)

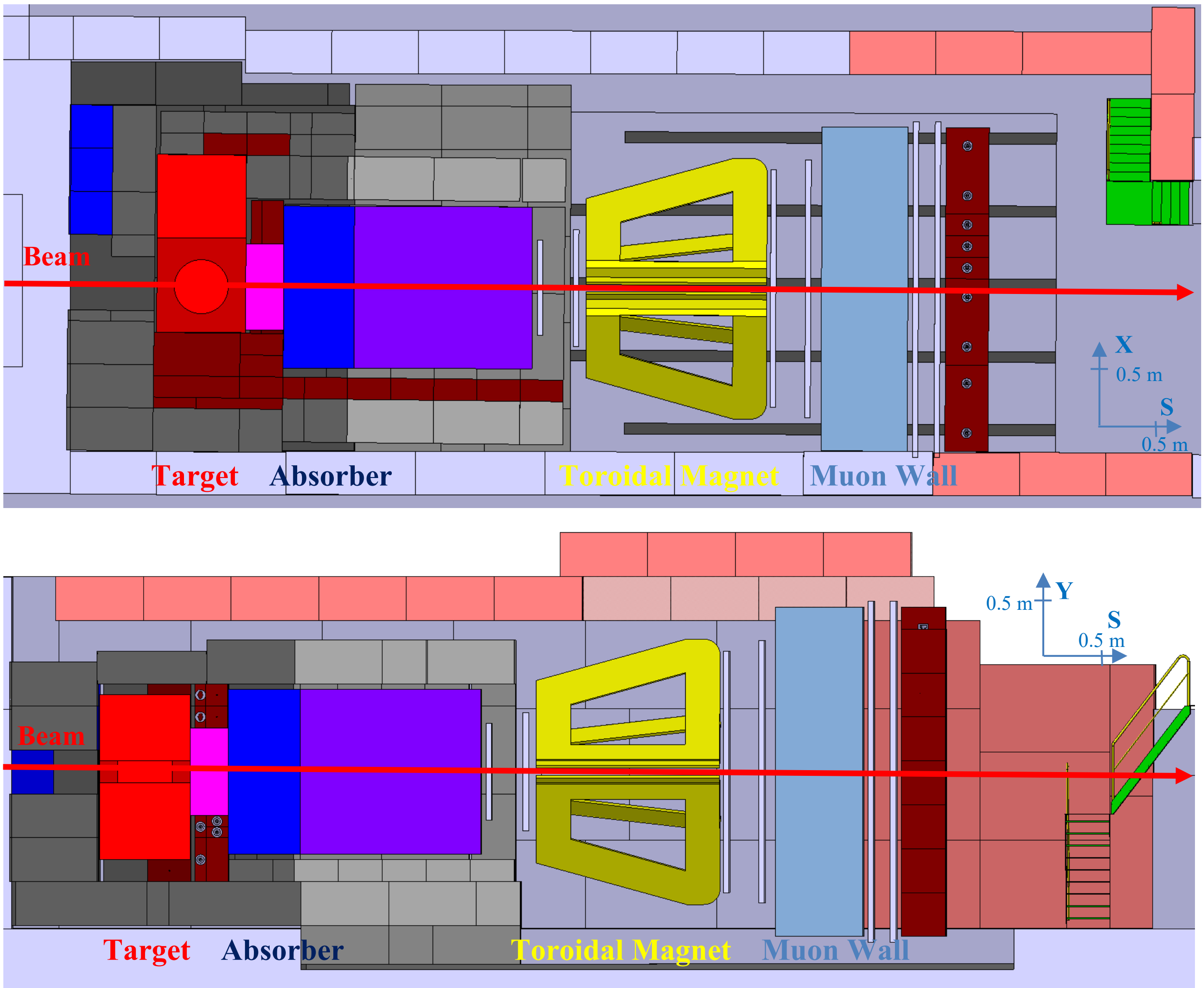

Figure 9: Drawing of the long (13.7 m) setup of NA60+ installed in the modified PPE1138 zone, top view (top) and side view (bottom). Compared with the short design presented in Fig.8 the detector has been moved downstream and the additional absorber (purple) has been added.

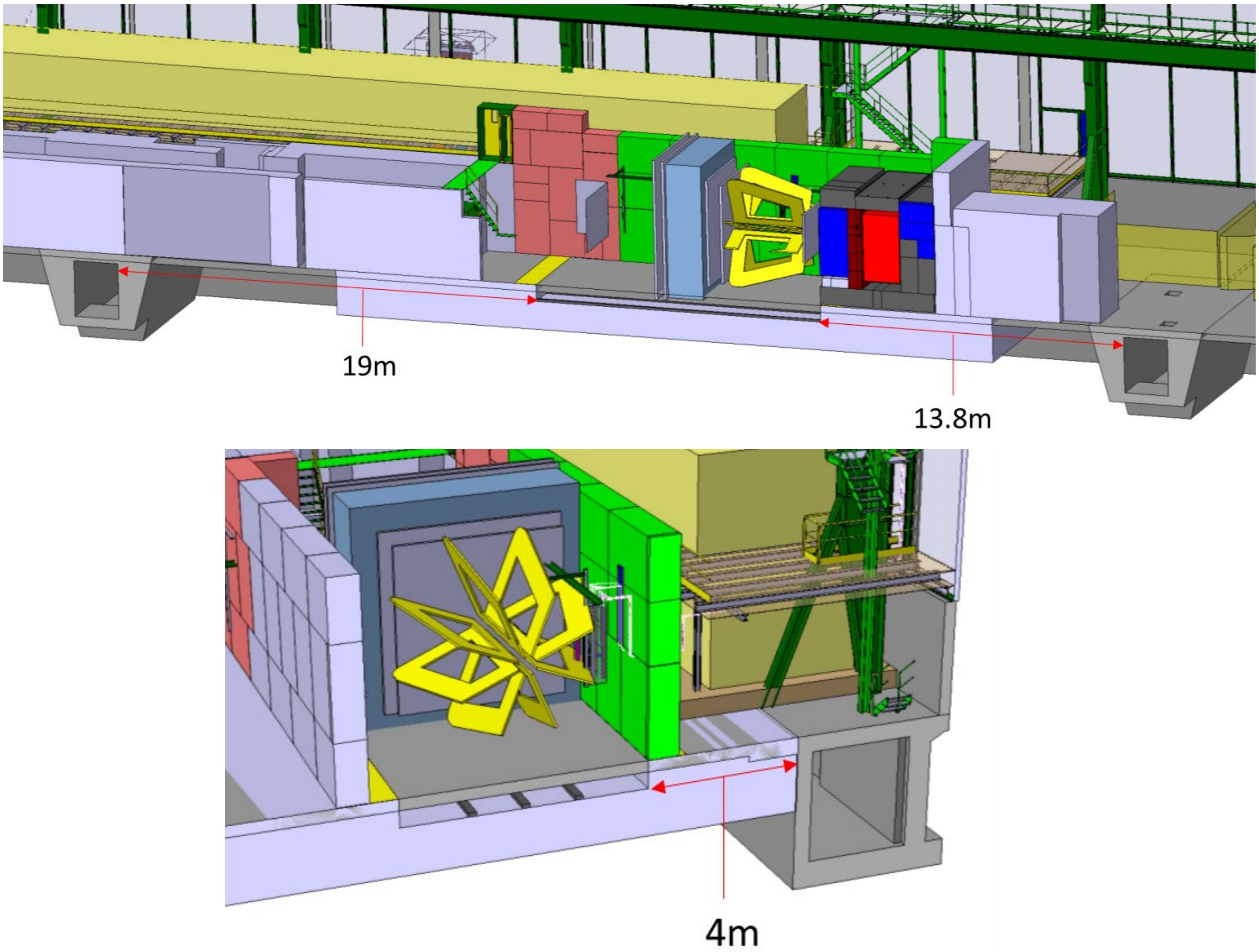

Figure 10: 3D drawings of the proposed installation of NA60+ in the zone PPE138 and the distance of the required excavation to the transverse (top) and longitudinal (bottom) underground galleries of EHN1 hall. Ther distances between the excavation edges and the closest location of the galleries are indicated.

The dipole magnet around the NA60+ target is shown in Figure 8 and Figure 9 in light red colour. The integration includes the installation of the additional shielding, required due to the radiation protection considerations described in Section 4. It includes the concrete and iron shielding blocks in the region around the target and behind the muon wall, marked as grey and dark-red blocks in Figure 8 and Figure 9, as well as the installation of the roof shielding, shown in Figure 11.

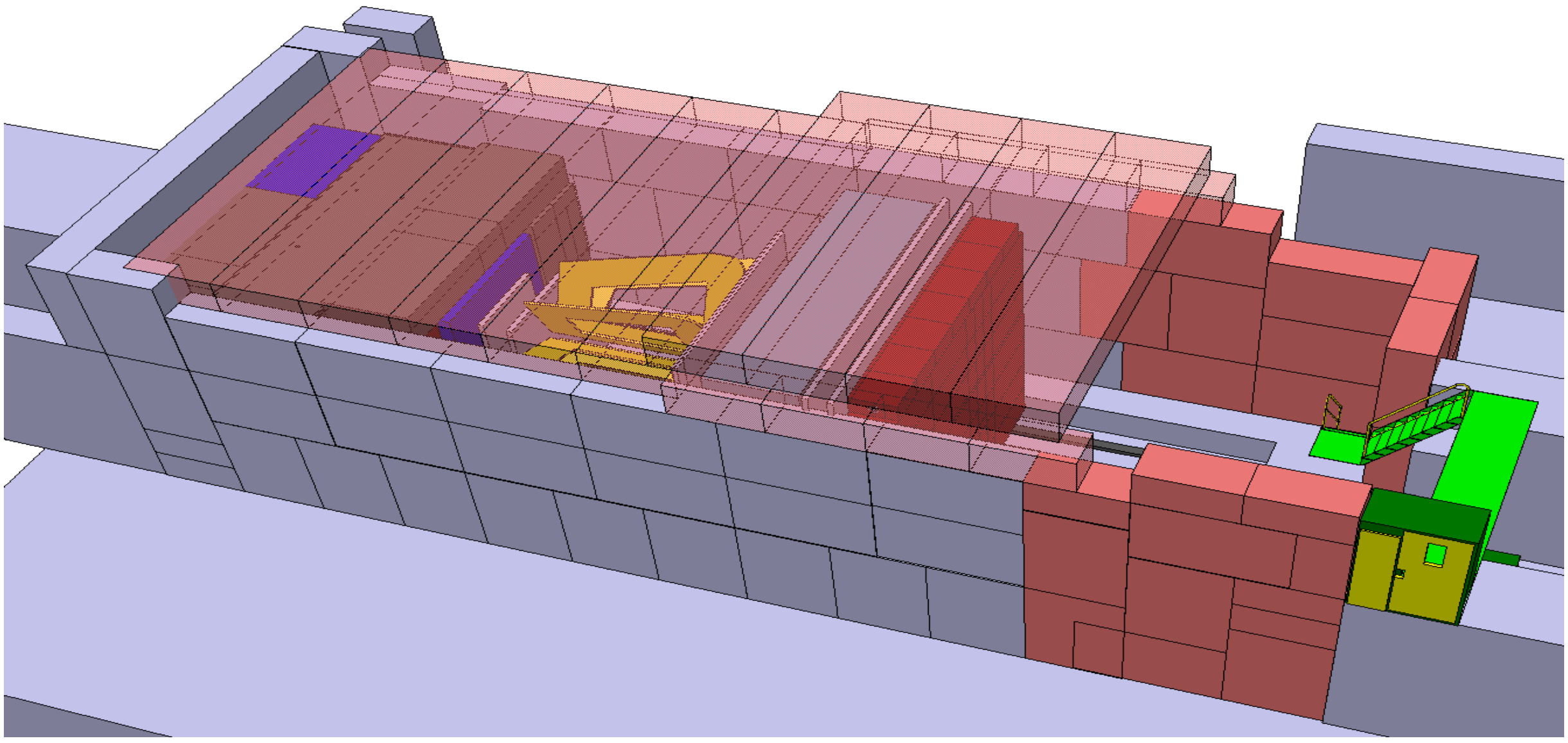

Figure 11: 3D drawing of zone PPE138 with NA60+ detector and the additional shielding.

## RADIATION PROTECTION STUDIES

Since NA60+ aims at pushing the beam intensity by at least one order of magnitude with what is currently delivered to EHN1, a detailed radiation protection assessment needed to be performed. The area surrounding the proposed location of NA60+ is classified according to CERN's radiological classification [11] as a Supervised Radiation Area with a low occupancy zone on one side (15 μSv/h limit) and permanent workplaces on the other (3 μSv/h limit). The shielding structure therefore has to be designed in such a way as to sufficiently reduce the prompt radiation to be compatible with the ambient dose equivalent rate limits linked to the area classification. The residual dose rates and air activation were also analysed, along with accidental beam losses in the beamline upstream of the experiment. The assessment was based on the FLUKA Monte Carlo particle transport code [6, 7]. FLUKA is a valuable tool used at CERN for various applications including radiation protection studies [12, 13] and benchmarking activation studies [14, 15, 16]. The given FLUKA simulations were performed using the latest released version (FLUKA 4-1.0), while the geometry was created using FLAIR [8].

### *Shielding Layout*

The proposed shielding layout for NA60+ for the case of 160 A GeV/c is depicted in Figure 12. In the most critical region around the target and the absorber a first layer of iron shielding, providing a higher attenuation of the radiation than concrete shielding, was implemented, which is then followed by additional concrete shielding. A chicane upstream of the target was added to allow access to the target region under certain conditions. A concrete shielding roof spanning the whole detector setup was added to reduce skyshine radiation.

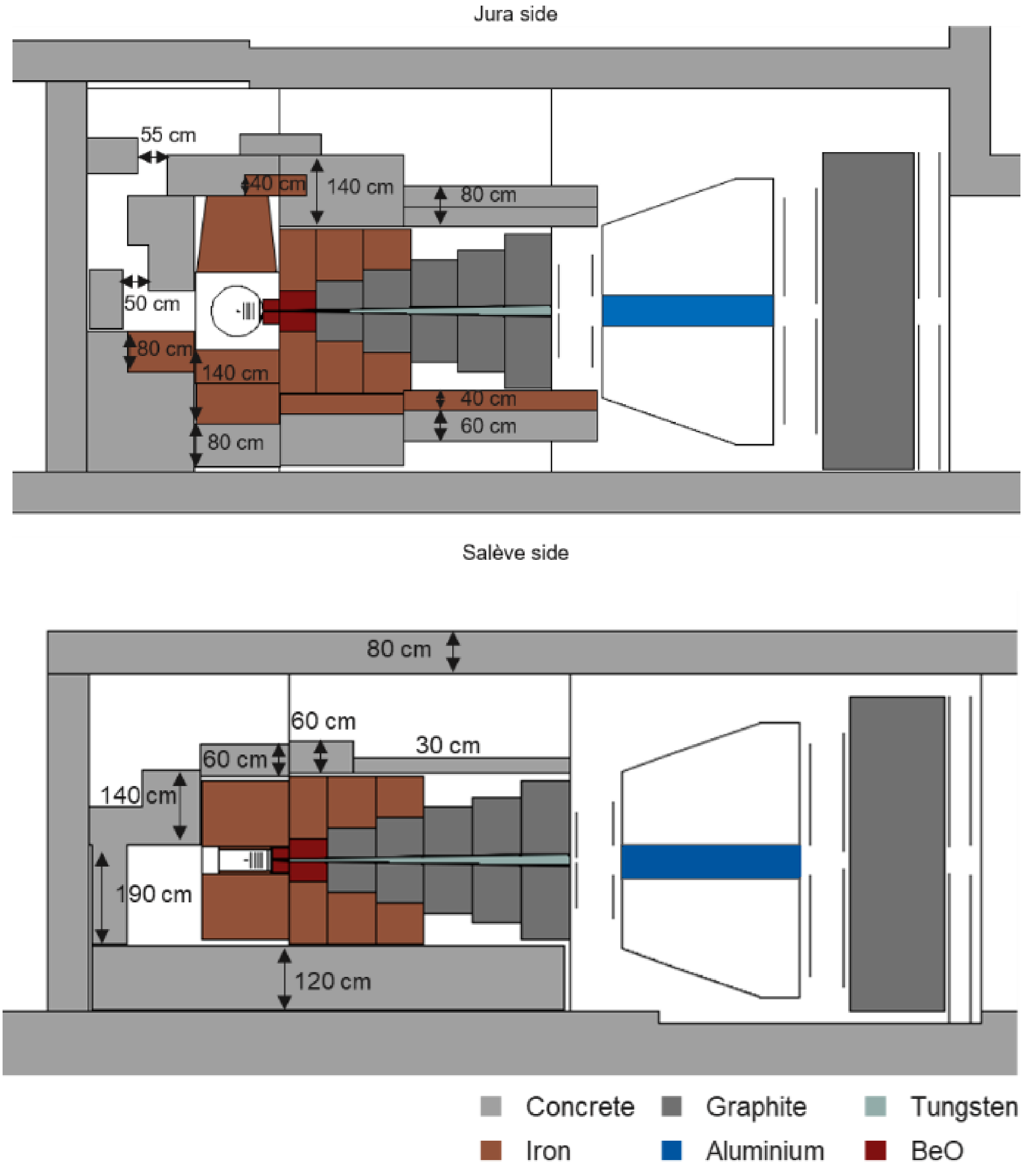

Figure 12: Shielding layout of NA60+ for 160A GeV/c as implemented in FLUKA with view from the top (top) and from the side (bottom) [5].

*Prompt dose rates*

Figure 13 and Figure 14 depict the prompt ambient dose equivalent rate distributions H*(10) for the experimental zone and its surroundings. The experimental zone PPE138, as all other experimental zones, is closed and not accessible by personnel during the operation and the personnel can be located only outside of the zones, e.g. in the corridor, crane or the counting rooms. It shall be noted that a safety factor 3 was taken into account for uncertainties related to material densities, geometry, beam parameters, simulations, etc. Hence, the displayed doses are 3 times greater than the simulation output for nominal intensity. The results show that the shielding allows sufficient reduction of the ambient dose rates to comply with the 3 μSv/h and 15 μSv/h dose rate limits. However, towards the top at the level of the crane driver cabin, which is located at a height of 7.65 m from the floor, the dose rate slightly exceeds the 15 μSv/h dose rate limit. During the beam operation with 160 A GeV/c lead ions a crane exclusion zone above the experiment will therefore need to be put in place.

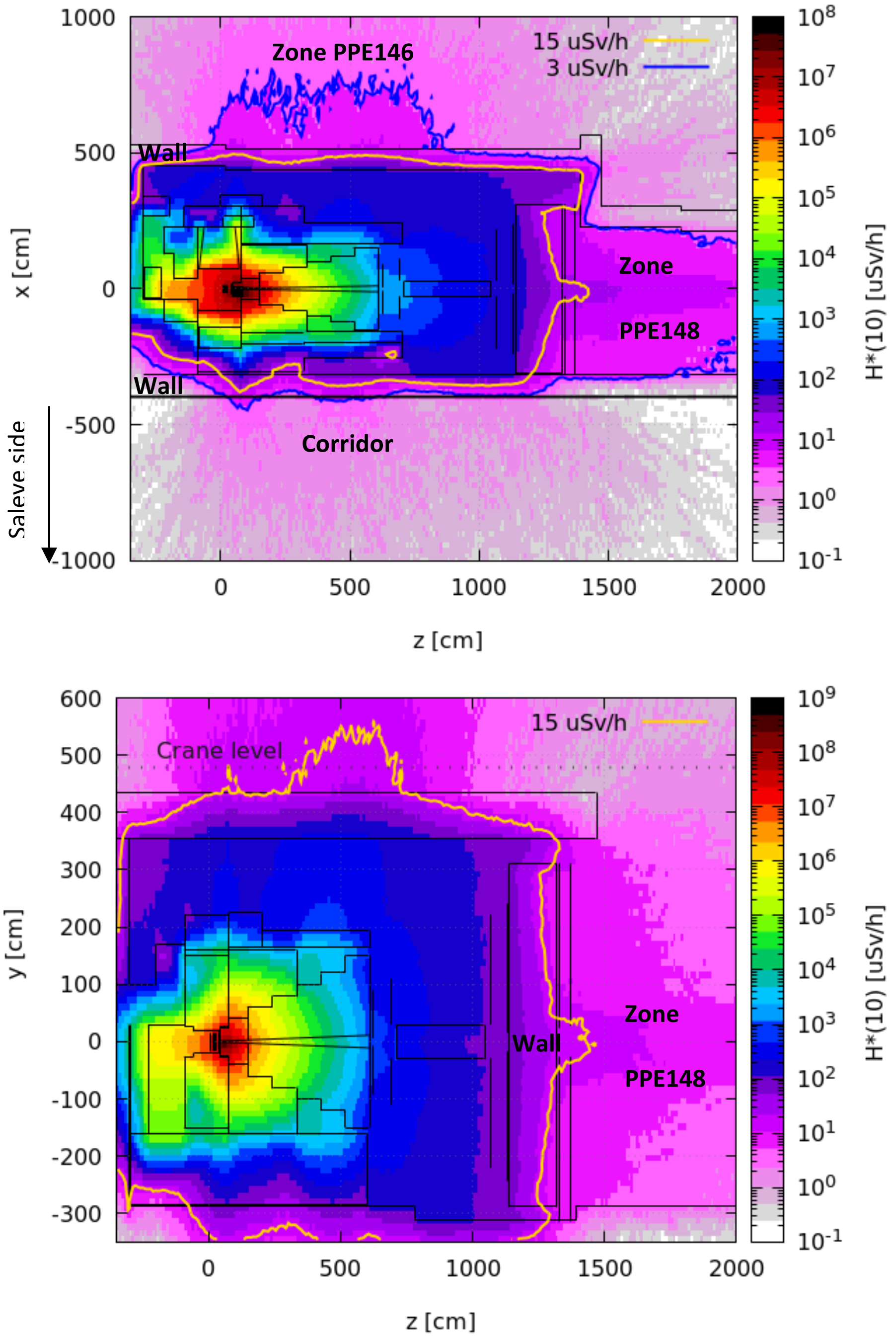

Figure 13: View from above (top) and side (bottom) of the prompt ambient dose equivalent rate in μSv/h for $1 \times 10^7$ Pb ions/spill of 160A GeV/c with 2 spills of 10 s every 40 s. The horizontal cut in the top figure is vertically averaged over ±50 cm around the beam axis. The vertical cut (Fig. 13) is horizontally averaged over ±40 cm around the beam axis. The red and blue lines illustrate the 3 μSv/h and 15 μSv/h dose rate limits for a Supervised Radiation Area with permanent and low occupancy workplaces, respectively.

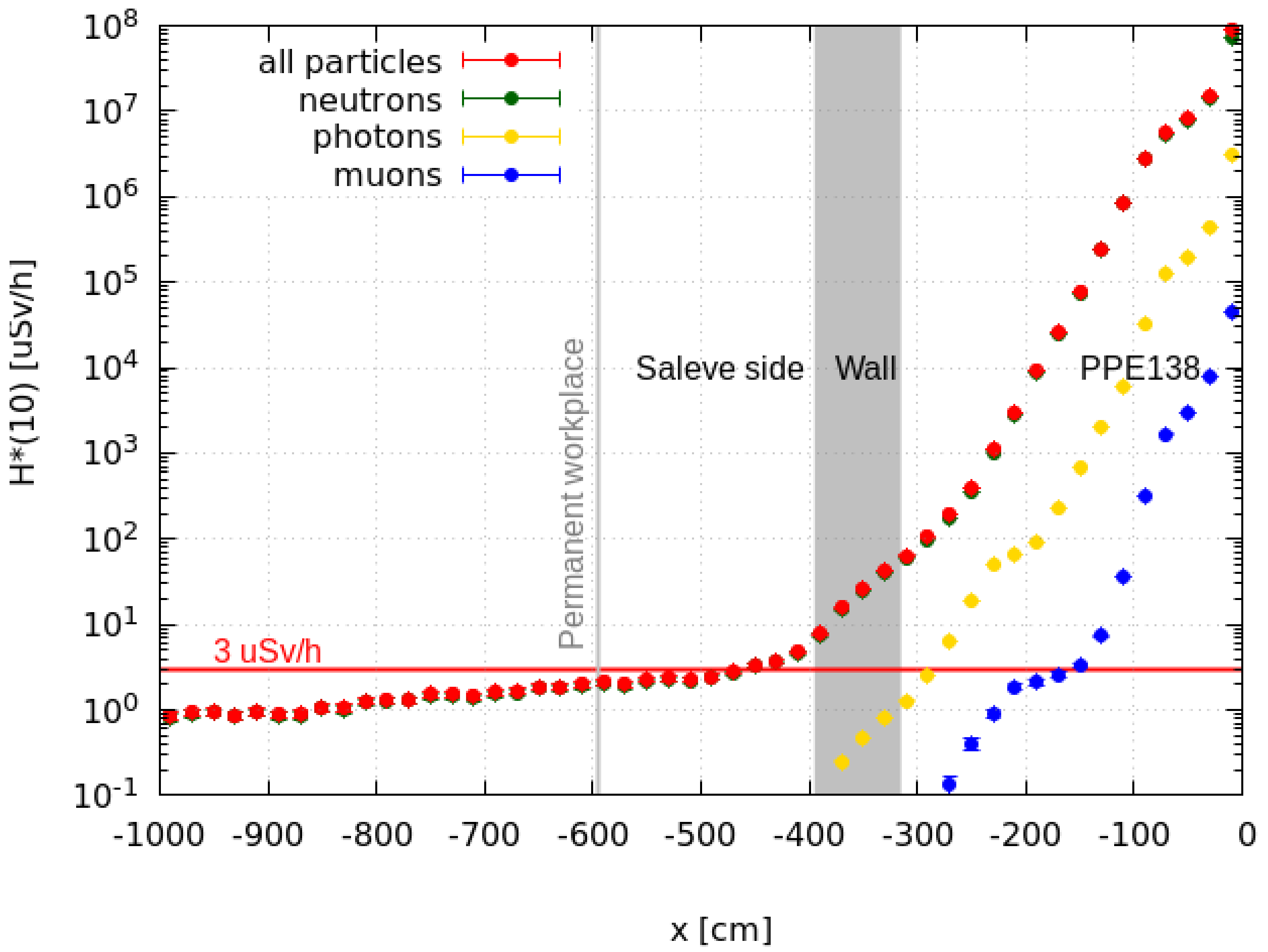

Figure 14: Vertical transverse cut of the prompt ambient dose equivalent rate (in µSv/h) for 1 $\times 10^7$ Pb ions/spill of 160A GeV/c with 2 spills each 40 s. The values are obtained by averaging vertically over ±50 cm around the beam axis and longitudinally over 50-100 cm behind the front of the target. The red line illustrates the 3 µSv/h dose rate limits for a Supervised Radiation Area with permanent workplaces.

*Residual Dose Rates*

The residual dose rates after 4 weeks of beam operation with 160 A GeV/c lead ions are presented in Figure 15 for different decay times. Also here a safety factor 3 was taken into account. The results show that the area outside of the shielding is compatible with a Supervised Radiation Area even for short decay times. However, close to the target the dose rates largely exceed the given 15 µSv/h limit of a Supervised Radiation Area. Directly upstream of the dipole magnet encompassing the NA60+ target, the dose rates reach approximately 440 µSv/h and 12 µSv/h after 1 minute and 1 week of cooling, respectively. That implies that one week of cooldown should be foreseen before general controlled access to the area is given.

In view of the high residual dose rates in the target area, any access to the chicane leading to the target area is foreseen to be regulated by a specialized procedure. Access will be granted only with the required training for work in such high radiation areas and under supervision by a representative from the CERN Radiation Protection Group (a condition not applicable for the majority of the user zones in EHN1). For shorter cooling times, where the ambient dose rates exceed the limit of a Simple Controlled Radiation Area (50 μSv/h limit), an operational dosimeter (DMC) is required next to the passive dosimeter (DIS). Furthermore, any work in the highly activated area must be optimized. Next to that, measures to prevent uncontrolled access to the area are to be foreseen.

*Air Activation*

To evaluate air activation, the particle fluences were scored in the air regions of experimental zone and then combined with the energy-dependent radionuclide production cross-sections using the ActiWiz Creator tool [17]. The dose due to inhalation of activated air was calculated by using the guidance value for airborne activity CA* and the inhalation dose coefficients $e_{inh}$ from the Swiss Radiological Protection Ordinance [18]. The dose was estimated conservatively with 4 weeks of beam operation at maximum intensity and with no air exchange. The intensity per year has been calculated based on the assumptions of $1 \times 10^7$ Pb ions/spill of 160A GeV/c with 2 spills each 40 s, 4 weeks (28 days) of operation per year, thus yielding 1.2E12 ions per year on the NA60+ target.

When assuming full mixing between the air regions and no cooling, the specific airborne radioactivity amounts to 0.02 CA and therefore lies below the given limit of 0.1 CA for a Supervised Radiation Area at CERN [11]. Here, the largest contribution comes from the short-lived radionuclides $^{41}Ar$, $^{13}N$, $^{15}O$ and $^{11}C$. The dose from inhalation during 1 hour of stay was estimated to be of 0.006 μSv, with the main contribution coming from $^{14}C$, $^{32}P$, $^{7}Be$, $^{33}P$ and $^{35}S$.

---

* Exposure to an airborne activity concentration of 1 CA for 40 hours per week and 50 weeks per year yields a committed effective dose of 20 mSv.

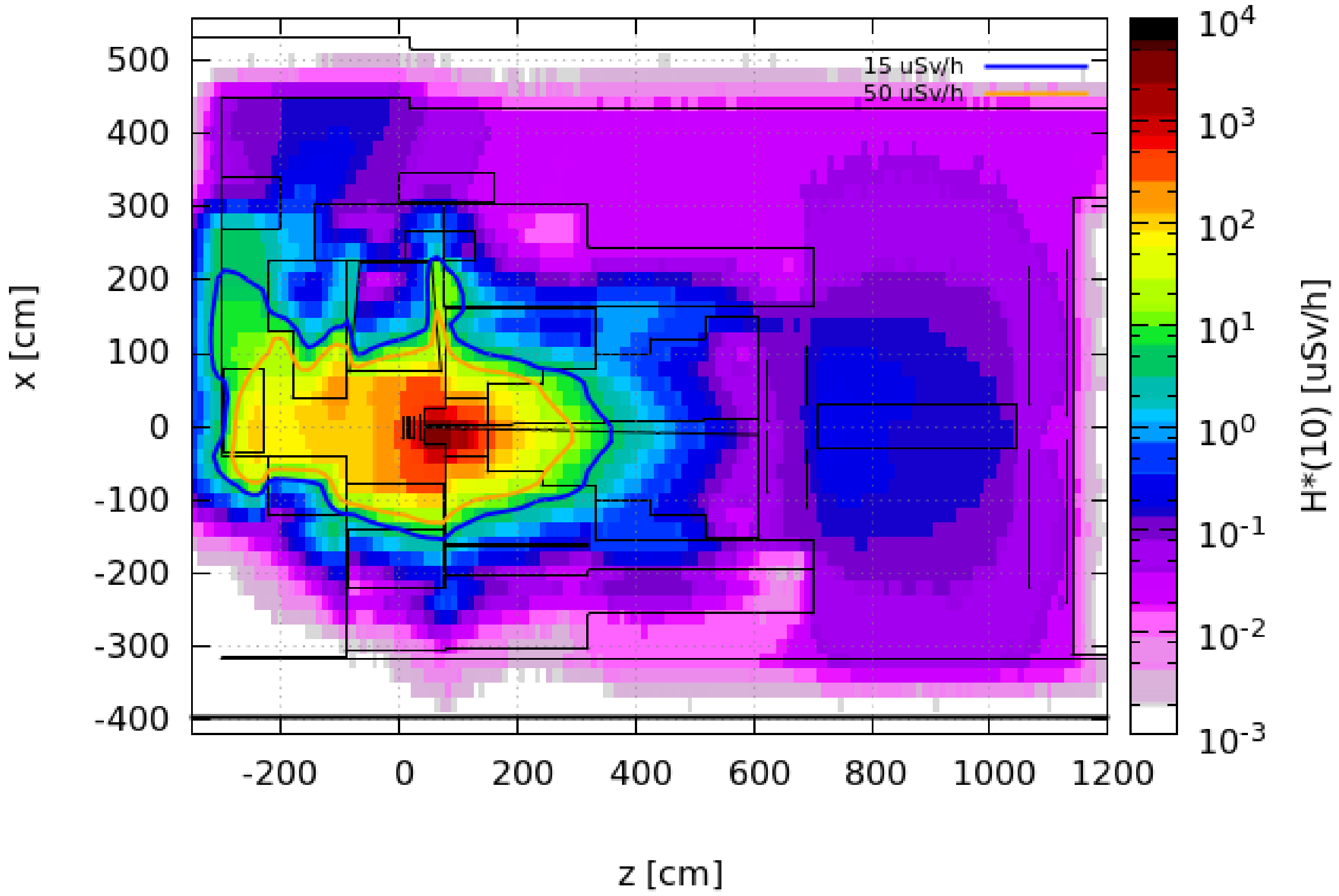

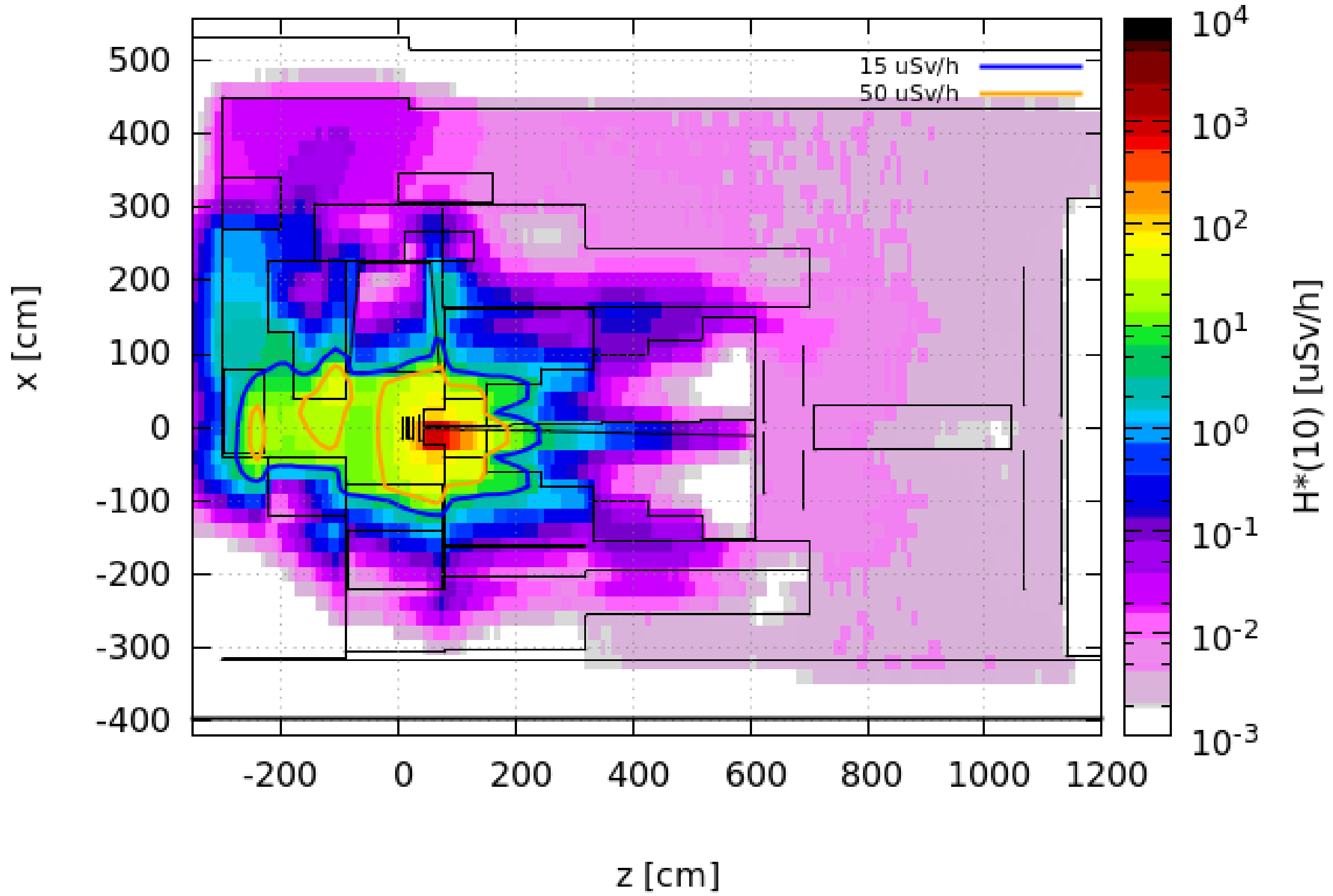

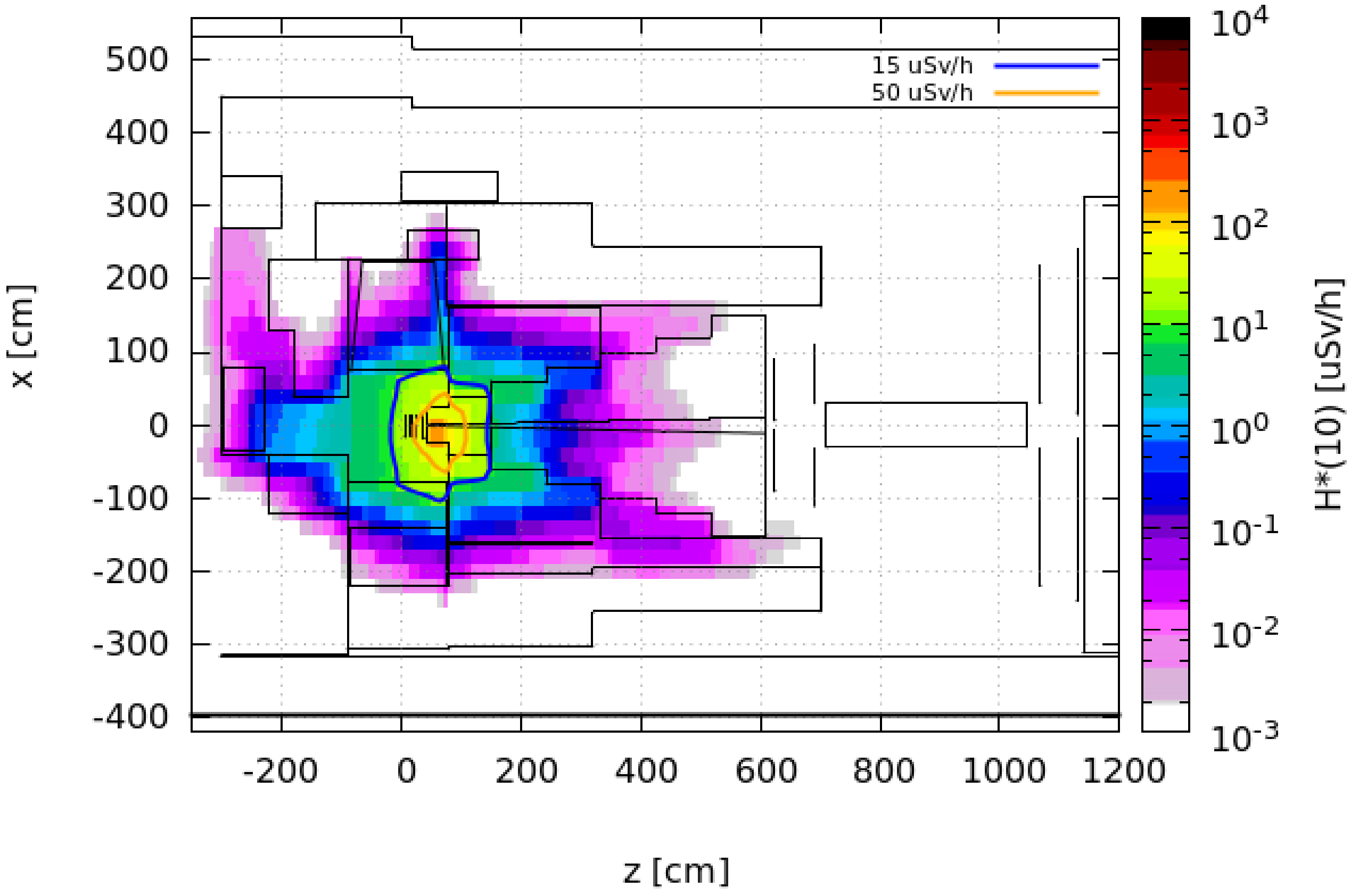

Figure 15: Top view of the residual ambient dose equivalent rate H*(10) (in µSv/h) for 1 minute (top), 1 day (middle) and 1 week (bottom) of decay after 4 weeks of operation 1.2 $\times 10^{12}$ ions o target. The horizontal cut is vertically averaged over ±50 cm around the beam axis. The blue and yellow lines illustrate the 15 µSv/h and 50 µSv/h dose rate limits for a Supervised and Simple Controlled Radiation Area, respectively.

*Accidental Beam Loss*

Protective measures need to be put in place to mitigate the impact of accidental beam loss upstream of the NA60+ experiment. The shielding layout of the upstream region exhibits at several locations only 80 cm of concrete shielding. To study the radiation levels for an accidental beam loss upstream of the experiment, it was assumed that the beam is lost on a massive object (iron cylinder of R = 30 cm, L = 200 cm), in a part of the zone where only 80 cm of concrete shielding is present. It resulted that the accidental loss of only one single spill can already cause a dose exceeding 15 µSv. At least 160 cm of concrete shielding in the upstream region is therefore required. Furthermore, shielding chicanes for the access doors need to be implemented, and a crane exclusion zone has to be established to prevent the crane to travel above the zone during NA60+ beam operation.

While there is an extensive radiation protection monitoring system covering the most critical areas of EHN1 [19, 20] which raises an alarm in case the radiation levels exceed the set limits, there are currently no dedicated monitors to detect beam losses upstream of the experiment. Additional monitors would therefore need to be installed in this zone.

## CONCLUSIONS

The proposed integration of the NA60+ experiment in the EHN1 surface experimental hall has been examined concerning beam physics requirements, the infrastructure integration and radiation protection. The experiment is deemed to be feasible with regard to these aspects. The detector design, data acquisition, analysis and physics reach will be the treated in a Letter of Intent currently in preparation.

## ACKNOWLEDGEMENTS

The authors would like to kindly thank numerous colleagues from NA60+ experiment for the fruitful discussions and their input, as well as Dr. Lau Gatignon and Dr. Nikolaos Charitonidis for their insights, recommendations and measurement results in regard to ion beams in CERN North Area as well as the integration aspects.